\documentclass[twocolumn]{aastex63}

\shorttitle{Ho IX X-1 Radio Bubble}
\shortauthors{Berghea et al. 2019}

\begin{document}

\title{Detection of a Radio Bubble around the Ultraluminous X-ray Source Holmberg IX X-1}

\email{tberghea@yahoo.com}

\author[0000-0001-5538-5903]{Ciprian T. Berghea}
\affiliation{U.S. Naval Observatory, 3450 Massachusetts Avenue NW, Washington, DC 20392, USA}

\author[0000-0002-4146-1618]{Megan C. Johnson }
\affiliation{U.S. Naval Observatory, 3450 Massachusetts Avenue NW, Washington, DC 20392, USA}

\author[0000-0002-4902-8077]{Nathan J. Secrest} 
\affiliation{U.S. Naval Observatory, 3450 Massachusetts Avenue NW, Washington, DC 20392, USA}

\author[0000-0003-0945-5610]{Rachel P. Dudik}
\affiliation{U.S. Naval Observatory, 3450 Massachusetts Avenue NW, Washington, DC 20392, USA}

\author[0000-0001-8017-5115]{Gregory S. Hennessy}
\affiliation{U.S. Naval Observatory, 3450 Massachusetts Avenue NW, Washington, DC 20392, USA}

\author{Aisha El-khatib}
\affiliation{U.S. Naval Observatory, 3450 Massachusetts Avenue NW, Washington, DC 20392, USA}

\begin{abstract}

We present C and X-band radio observations of the famous utraluminous X-ray source (ULX) Holmberg IX X-1, previously discovered to be associated with an optical emission line nebula several hundred pc in extent. Our recent infrared study of the ULX suggested that a jet could be responsible for the infrared excess detected at the ULX position. The new radio observations, performed using the Karl G.\ Jansky Very Large Array (VLA) in B-configuration, reveal the presence of a radio counterpart to the nebula with a spectral slope of -0.56 similar to other ULXs. Importantly, we find no evidence for an unresolved radio source associated with the ULX itself, and we set an upper limit on any 5~GHz radio core emission of 6.6 $\mu$Jy ($4.1\times10^{32}$ erg s$^{-1}$). This is 20 times fainter than what we expect if the bubble is energized by a jet. If a jet exists its core component is unlikely to be responsible for the infrared excess unless it is variable. Strong winds which are expected in super-Eddington sources could also play an important role in inflating the radio bubble. We discuss possible interpretations of the radio/optical bubble and we prefer the jet+winds-blown bubble scenario similar to the microquasar SS~433.

\end{abstract}


\section{Introduction} \label{section: Introduction}

Ultraluminous X-ray sources (ULXs) are unusually bright off nuclear X-ray sources detected in nearby galaxies, with L$_X$ $>$ $10^{39}$ erg s$^{-1}$, which is approximately the Eddington limit for a 10~M$_{\odot}$ accretor. After ULXs were discovered roughly three decades ago, it was suggested they may harbor intermediate-mass black holes \citep[IMBH, e.g.][]{col99, mil04} with $M_\mathrm{BH}\gtrsim100$~$M_\sun$. This interpretation was mostly based on the so-called cool accretion disk component found by modeling the X-ray spectra, as such cool disks were theoretically expected in accreting IMBHs \citep{mil05}. Later studies using more and better quality data  have shown that this cannot be taken as evidence for IMBHs as many other explanations are possible for the soft component, which does not necessarily imply cool disks \citep{ber08, glad09}. In particular in the  super-Eddington accretion we also expect a soft component, which is interpreted in this case as coming from the inner photosphere of the outflowing wind \citep[e. g.][]{pou07}.

After years of multiwavelength observations, very few ULXs remain strong candidates for the IMBH scenario, such as HLX-1 in the galaxy ESO 243-49 \citep{farr09, webb12} and M82 X-1 \citep{str03}. The majority of ULXs are now thought to be super-Eddington accreting stellar-mass compact objects possibly accompanied by mild beaming \citep{king01}. In the last few years several of the known ULXs were actually shown to contain neutron stars. For more information on ULXs see the recent review of \citet{kaa17}.

The ULX Holmberg IX X-1 (hereafter Ho IX X-1) has been well studied in the X-rays and optical. One of the most notable features of this ULX is a large nebula spanning a region $\sim300\times400$~pc in extent. The ultimate power source for the nebula, like similar structures around other ULXs, has been a matter of debate \citep[e. g.][]{moon11}.

The presence of these nebulae around many ULXs and the fact that there is little evidence for IMBHs suggests that most are similar to the famous super/hyper-Eddington microquasar SS~433 as it was suggested early on by many authors \citep{fab01, fab06, beg06}. Even HLX-1, which is still used as the best candidate for IMBHs could actually be such a system \citep{king14}. 

Most such nebulae were found in the optical and only a handful have been detected in the radio: Ho~II~X-1, IC~342~X-1, NGC~5408~X-1,  S26 in NGC~7793 and MF~16 in NGC~6946.
There are also a few microquasars in our Galaxy and the Local Group that have associated radio bubbles besides SS~433/W50: IC~10~X-1  \citep{hee15}, and also Cygnus X-1 \citep{gal05} and Circinus X-1 \citep{hei13}. A more recent discovery is S10 in NGC~300 \citep{urq19}.

For Ho IX X-1, most indications based on optical studies (He\,\textsc{ii}/H$\beta$ ratio) show that the nebula is produced by shocks as opposed to photoionization \citep[e. g.][]{abo07, ber10}, suggesting jet activity similar to the W50 bubble associated with SS~433. However,  more recent observations show that photoionization is also present \citep{moon11, abo08} similar to other ULX nebulae.

In our previous work on Ho~IX X-1 \citep{dud16}, we found an intriguing IRAC infrared excess suggesting the presence of a possible jet. We  proposed for deep VLA observations, down to 5~$\mu$Jy, in the C and X bands to test this hypothesis. In this work, we discuss the results of this campaign. We present our radio observations. We discuss the physical properties of the bubble and possible scenarios for its origin in Section~\ref{section: Discussion}, and we give our conclusions in Section~\ref{section: Conclusions}.

\section{VLA Data} \label{section: VLA Data}
\subsection{Observations} \label{subsection: Observations}

We obtained C- and X-band Karl G.\ Jansky Very Large Array (VLA) observations of Ho~IX~X-1 in B configuration (Project 16A-065; PI: R.\ Dudik) with a total time allocation of 3.3 hours. The data were calibrated using \textsc{casa} version 4.7.0 and imaged with version 5.3.0. 
We began each observing session by integrating for six minutes on the primary calibrator 3C~286.  We then alternated between our target source and a secondary calibrator, J1048+7143, by interweaving two minute integrations of the secondary calibrator for every 10 minutes of integration time on the target source.  In this way, we were able to obtain accurate phase calibrations over the duration of our observations.

\subsection{Calibration} \label{subsection: Calibration}
\subsubsection{C-band}

The C-band data were obtained on 2016~May~26 and had significant radio frequency interference (RFI).  Therefore, the data were calibrated and flagged by hand using standard \textsc{casa} tasks.  The methodology used in the C-band calibration was modeled after a recipe used by the Continuum HAlos in Nearby Galaxies; an EVLA Survey (CHANG-ES) collaboration \citep[see][]{2015AJ....150...81W}.  An outline of the calibration steps is given here.  

The calibration began by applying antenna position corrections followed by antenna-based delay calibrations. Next, we Hanning smoothed our data to alleviate any ringing across the band due to strong RFI.  The data were then flagged by hand using the \textsc{casa} task, PLOTMS.  We determined that three spectral windows needed to be  flagged altogether to remove strong RFI, while the remaining 29 spectral windows were flagged one by one.  The absolute flux density scale for the total intensity, Stokes I, was determined using the primary flux calibrator, 3C~286, and applying the Perley$-$Butler 2013 flux density scale.  We used this primary calibrator to create an initial gain calibration table.  

We fit for the bandpass on the primary calibrator while correcting for the gains in real time using the initial gain calibration table.  We then determined gain solutions for the phase only for the primary calibrator, while fitting for the bandpass {\bf in real time} using the bandpass table created in the previous step.  Gain solutions for the phase were then determined for the secondary calibrator while fitting for the bandpass in real time in the same manner as for the primary calibrator.  After the gain solutions were determined for the phase, the amplitude gain solutions were derived for the primary flux calibrator and then for the secondary calibrator.

The calibrated fluxes from the primary calibrator were then bootstrapped to the secondary calibrator.  The final calibration tables were then applied to the primary calibrator, then the secondary calibrator, and then the source.  The results for both calibrators and the source were checked using PLOTMS.  If additional flagging was necessary, the calibration steps were repeated.  This process was iterated until all RFI was removed and the data were satisfactorily calibrated.  The data were then split so that each target was in a separate measurement set.

\subsubsection{X-band}

The X-band data were obtained on 2016~June~5 and had little RFI.  The data were again calibrated by hand using the same procedure as used for the C-band.  The absolute flux density scale was again determined using the primary flux calibrator, 3C~286 and applying the \citet{per13} flux density scale.

\subsection{Imaging}\label{sec:imaging}

Once the C- and X-band data were successfully calibrated, we used the \textsc{casa} version 5.3.0 task CLEAN to produce images.  The imaging process was a time consuming step as multiple iterations were required to produce high fidelity maps.  

\begin{figure}
\includegraphics[width=\columnwidth]{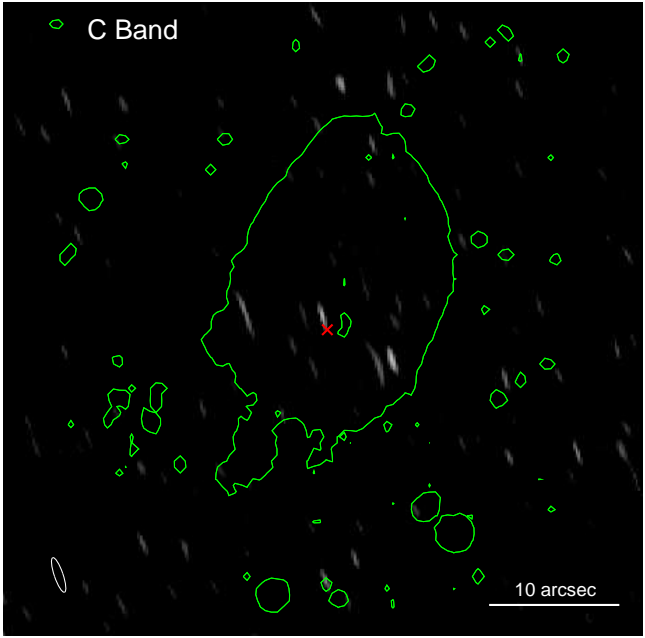}
\caption{C-band continuum with Subaru H$\alpha$ contours  overlaid. The position of the ULX is marked with a red 'x'. The faint source just north of the ULX is not significant at the 3$\sigma$ level and has an upper limit of 6.6~$\mu$Jy, which corresponds to a luminosity of  $4.1\times10^{32}$ erg s$^{-1}$.}
\label{point}
\end{figure}

We began with the calibrated C-band data and produced a ``dirty'' image of the FWHM of the primary beam to determine if there were strong continuum sources in the field.  There were no continuum sources that appeared to cause any obvious artifacts that would impede the imaging process; however, the dirty image showed hints of low surface brightness extended, possibly diffuse emission at the phase center coincident with the ULX.  Nonetheless, we began our imaging with a ``standard'' clean using a robust=0.0 weighting and the ``clarkstokes'' point spread function (psf) mode and cleaned down to an rms threshold of 3$\sigma$ where we used a theoretical rms noise of $\sigma$ = 2.2 $\mu$Jy beam$^{-1}$ based on the integration time.  Due to the high decl. of the source and the relatively short on-source integration time of $\sim$75 minutes for the C-band, the synthesized beam of the robust=0.0 weighting used in the initial standard clean was extremely elongated and produced an FWHM beam of 2$\farcs$8 $\times$ 0$\farcs$7 at a position angle of 19$\fdg$8.
This initial image (not shown) showed no emission at the phase center and therefore, we began exploring more complex imaging methods.  

\begin{figure*}
\includegraphics[scale=0.5]{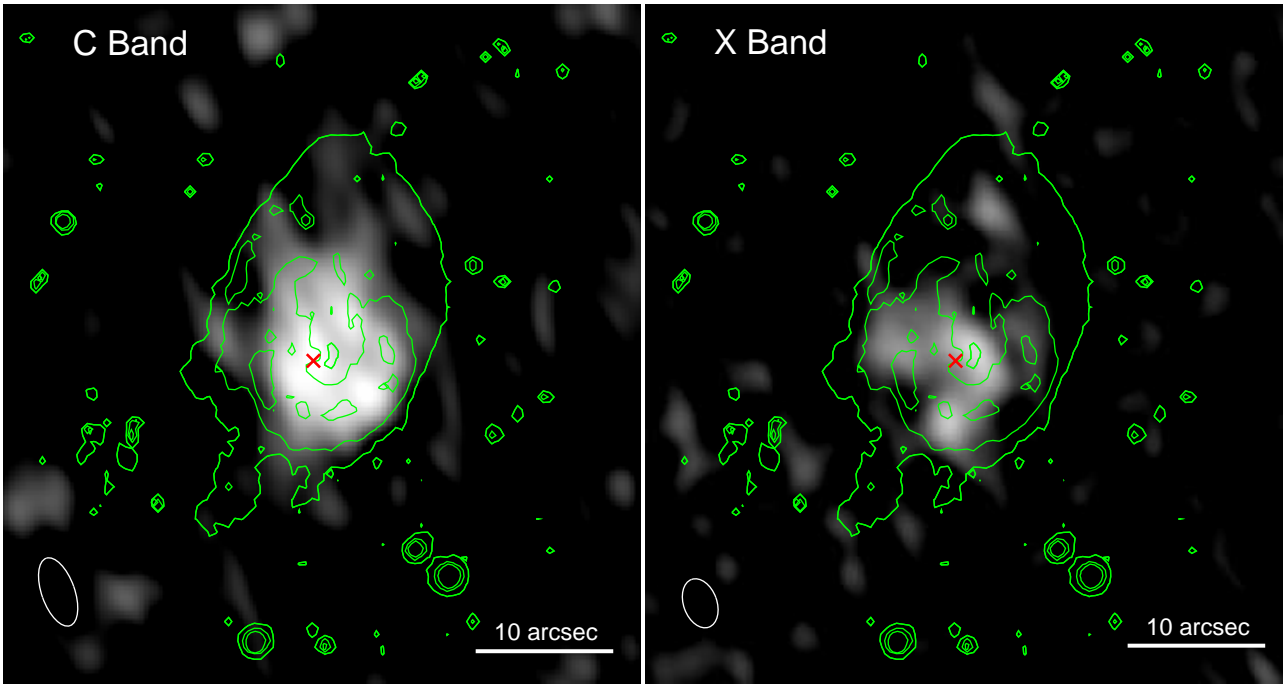}
\caption{C- (left) and X-band (right) continuum images using a $uv$-taper of 50,000$\lambda$ with H$\alpha$ contours overlaid. The position of the ULX is marked with a red 'x'. At the assumed distance of 3.6 Mpc, 10 arcsec is about 175~pc.}
\label{contours}
\end{figure*}

We implemented the multiscale, multifrequency synthesis (ms-mfs) algorithm \citep{Rau & Cornwell 2011} with the number of Taylor series expansion terms (nterms) set to two with a natural weighting and four spatial scales equal to 0 (``standard clean''), 2$\times$beam, 4$\times$beam, and 10$\times$beam. 
In this manner, we aimed to fit for the in-band spectral index, $\alpha$, assuming 
\begin{equation}
{I_{\nu}}\ \propto\ \nu^{-\alpha},
\label{eqn}
\end{equation}
where $I_{\nu}$ is the intensity at frequency $\nu$. This image produced two point-like detections near the phase center; however, the signal-to-noise was below 3$\sigma$ (see Figure \ref{point}).  Therefore, we re-imaged using the ms-mfs algorithm with ``Briggs'' robust = 2.0 weighting and a $uv-$taper of 120,000$\lambda$. With these parameters set, the resulting image began to show a diffuse, extended structure coincident with the location of the ULX.  The $uv-$tapering was reduced systematically until the final image shown in the left panel of Figure \ref{contours} was produced.  This image has a $uv-$taper of 50,000$\lambda$ and a uniform RMS noise of 5.3 $\mu$Jy beam$^{-1}$. We corrected the image for the drop-off in the primary beam and integrated the emission for a flux density measurement of 0.24$\pm$0.025 mJy. The FWHM of the beam is 5$\farcs$2 $\times$ 2$\farcs$4 at a position angle of 19$\fdg$5

We applied a similar approach to imaging the X-band data.  However, the resulting source structure appears dramatically different.  The total integrated flux density for the primary beam corrected X-band emission is 0.18$\pm$0.018 mJy and the uncorrected primary beam image with uniform noise has an RMS noise of 4.2 $\mu$Jy beam$^{-1}$. The FWHM of the beam is beam 3$\farcs$7 $\times$ 2$\farcs$4 at a position angle of 20$\fdg$3.

\begin{figure*}
\includegraphics[scale=0.4]{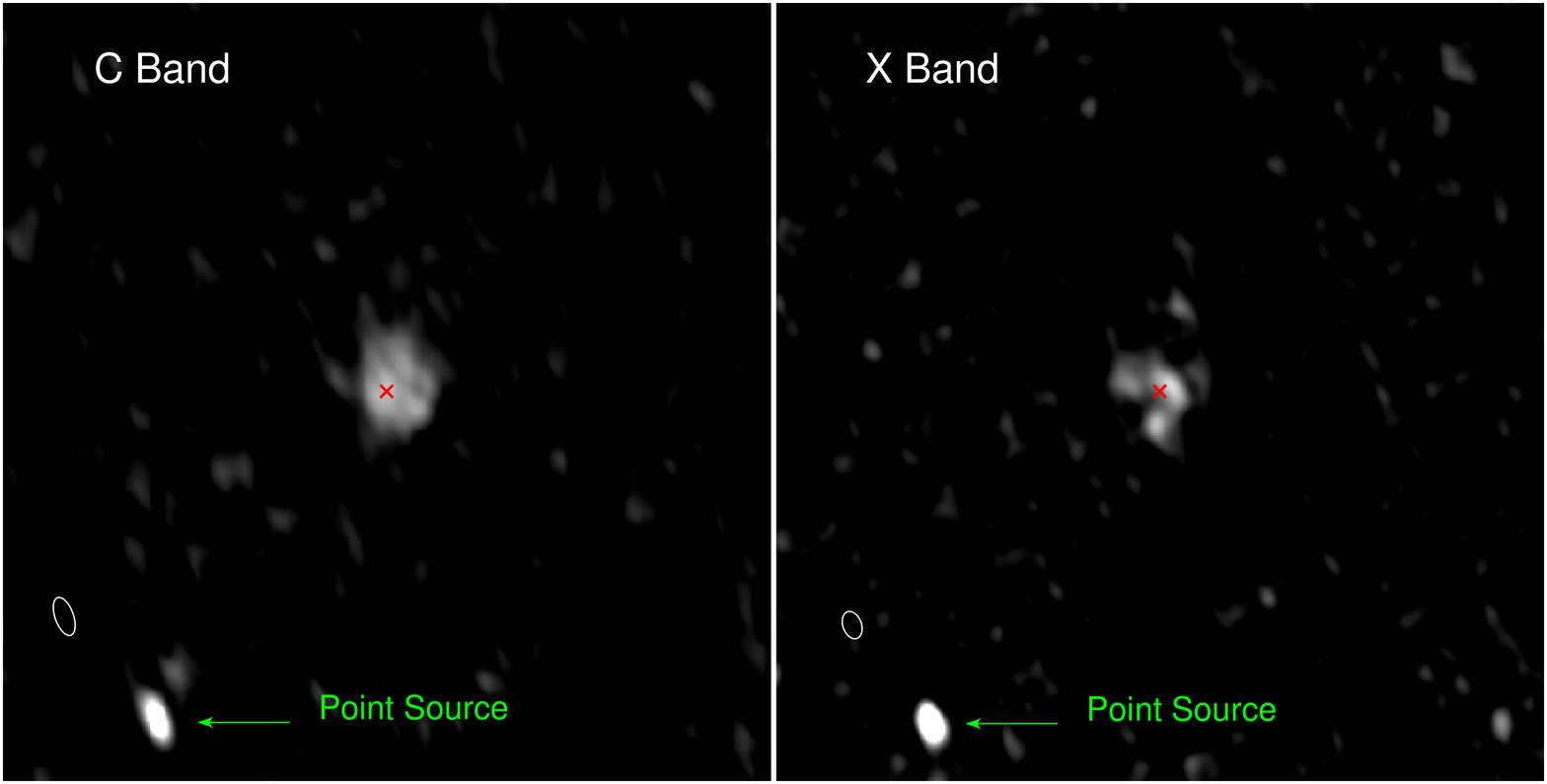}
\caption{C- (left) and X-band (right) continuum images showing a point source south of the ULX. This source has been used in Section \ref{sec:imaging} to asses the validity of the X-band structure. This image also shows that there is no evidence in either band of residual RFI or other spurious bad data.}
\label{point1}
\end{figure*}

To assess the validity of the X-band structure, we used a point source that lies to the south of the ULX and is visible in both bands as show in in Figure \ref{point1}. This image also shows that there is no evidence in either band of residual RFI or other spurious bad data. If there were problems with the calibration or imaging of the data at X-band, then, the point source would show a deviation in its point-like structure. We compare the peak to the integrated flux for the point source as a measure and test of the compactness of the source using the primary beam corrected intensity map for each band.  The C-band and X-band peak-to-integrated flux density ratios are 0.92 $\pm$ 0.12 and 1.01 $\pm$ 0.076, respectively, both of which are consistent with one and indicative of a perfect point source.  The peak-to-rms ratios for the C- and X-bands are 17 and 27, respectively, with the X-band having nearly a factor of 1.5 better signal-to-noise.  Due to the lack of any visual artifacts in the maps and the ratios of peak-to-integrated and peak-to-rms values being comparable between the bands, we conclude that the C- and X-band images are of similar quality both in calibration and image fidelity.

It is interesting that the structures between the two bands are so strikingly different.  We note that the intensity distribution between the bands near the center of the H$\alpha$ bubble is more prevalent in the X-band and more diffuse in the C-band.  The C-band radio emission appears most intense at the edge of the H$\alpha$ shell while the X-band emission appears to drop off.  The largest angular scale at X-band for the B-array configuration is $\sim$17$\arcsec$, which is roughly the major axis of the H$\alpha$ outer green contour.  Likewise, the largest angular scale for C-band at B-array is $\sim$29$\arcsec$.  Thus, it is probable that we have recovered most, if not all, of the emission interior to the green contours shown in Figure \ref{contours} at both frequencies.  However, we note that our experimental design was maximized for detection of a point source and not a diffuse extended ``bubble-like'' structure so longer $uv$-tracks across larger parallactic angles might aide in recovery of the extended emission especially if these observations occur at a more compact configuration.  However, by applying a uniform $uv$-taper across both frequency bands, we aimed to find a balance between matching resolution and maintaining sensitivity so that we could feasibly compare the integrated fluxes and draw a few first-order conclusions.

As Figure \ref{contours} shows, the radio bubble is slightly smaller in size than the optical counterpart, approximately 200 by 300 pc. We assume the same distance of 3.6~Mpc to Holmberg~IX as we used in \citet{dud16}, at this distance 1$\arcsec =$ 17.5~pc. We also note the large differences between the C and X band images. While the emission is in general located in a similar area around the ULX position, the X-band emission is less extended and shows a very different structure compared to the C-band.

Unfortunately, due to the diffuse, extended nature of the source and low signal-to-noise, an in-band spectral index was not possible to produce in either band for these observations.  

\section{Discussion} \label{section: Discussion}

\begin{figure*}
\centering
\includegraphics[width=0.9\textwidth]{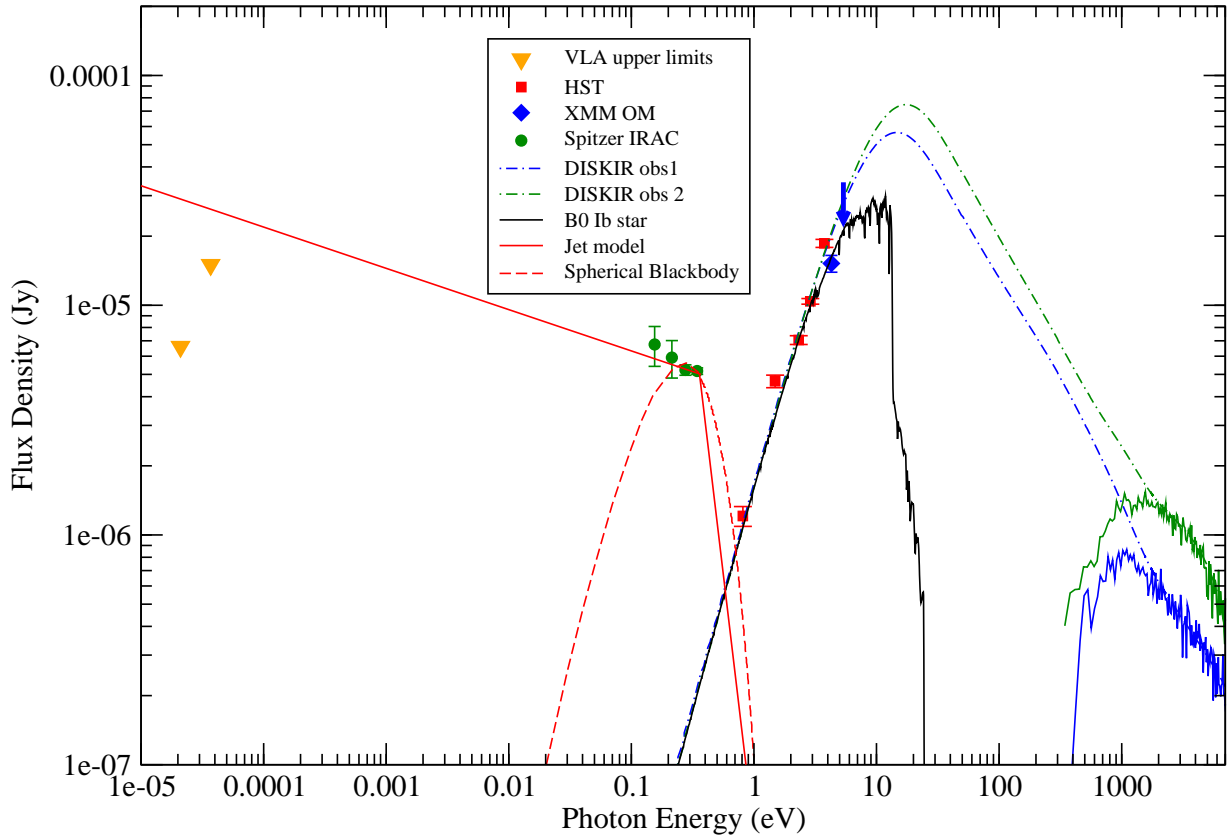}
\caption{SED from \citet{dud16} with upper limits for a radio point source measured from our VLA observations. We show the infrared excess we detected in that paper with extrapolated power-law into radio. Accretion disk models shown (DISKIR) are based on X-ray, UV and optical data. The optical/UV data alone can also be well fit by a companion B0~Ib star.
}
\label{sed}
\end{figure*}

The main motivation for our VLA observation was to search for a possible jet in Ho~IX X-1, based on our previous paper \citep{dud16}, where we found an IR excess and we suggested it could be caused by a jet. While there is a faint source very close to the ULX position in the first C-band image (see note in Figure~$\ref{point}$ caption), it is below the 3$\sigma$ upper limit of 6.6~$\mu$Jy. The upper limit on any ULX counterpart source in the X-band image is 21~$\mu$Jy, given the rms depth of the image. We show these upper limits in Figure~$\ref{sed}$ on the spectral energy distribution (SED) plot from \citet{dud16}. The C- and X-band upper limits are both below the level expected based on extrapolation from the Spitzer IRAC data. This does not rule out a jet because it may well be variable as seen for example in Ho~II X-1 \citep{cseh15} and SS~433 \citep{bro18}.

As stated in Section \ref{sec:imaging}, an in-band spectral index was not feasible due to the diffuse nature of the source. Furthermore, it was not possible to create a spectral index map of the source using the C- and X-bands as the source structure between the bands is so different. However, because of the uniform uv-taper that we applied to both the C- and X-bands, which provides comparable resolution between the bands, we found it appropriate to compare the integrated flux densities of the source between these bands and applied a standard power low using the commonly used spectral index definition provided in Equation \ref{eqn}.  In this way, we aim to assess to first order what the overarching emission mechanism is, i.e., thermal versus non-thermal, that is responsible for the continuum bubble that we observe.

The band-to-band spectral index of -0.56, is very similar to other radio ULX bubbles (Ho II X-1, IC 342 X-1, S26, MF~16 and IC~10) and suggests synchrotron emission. If we assume the radio emission is synchrotron we can calculate the minimum total energy contained in the nebula for equipartition following \citet{lon94}.  
We use the  radio luminosity density at 6 GHz, $L_{\nu} = 3.8\times10^{17}$~W~Hz$^{-1}$, the volume of the nebula  $5.6\times10^{56}$~m$^{3}$ and the fraction of the energy in protons to electrons $\eta$ - 1 = 0. We assume no protons as in \citet{fen99}, but this could be as high as 1000 and the minimum energy could then increase by a factor of 50 \citep[e. g.][]{cseh15}. On the other hand, the filling factor of the radio emitting material could be small and therefore decrease the energy requirement, for example, a filling factor of 0.1 will change the minimum energy by a factor of 0.37. Given these assumptions we obtain E$_{min} = 4.3\times10^{50}$ erg. This value is similar to other ULX radio nebulae and much higher than SS~433 (5$\times10^{45}$~erg). We can also calculate the corresponding magnetic field and the lifetime of the electrons, B$_{min}$ = 2.9 $\mu$G,  $\tau$ = 70.4 Myr.

Next we discuss two possible scenarios for the radio/optical bubble.

\subsection{Scenario 1: Super/Hypernova Remnant (SNR)}  

\begin{figure*}
\includegraphics[scale=0.6]{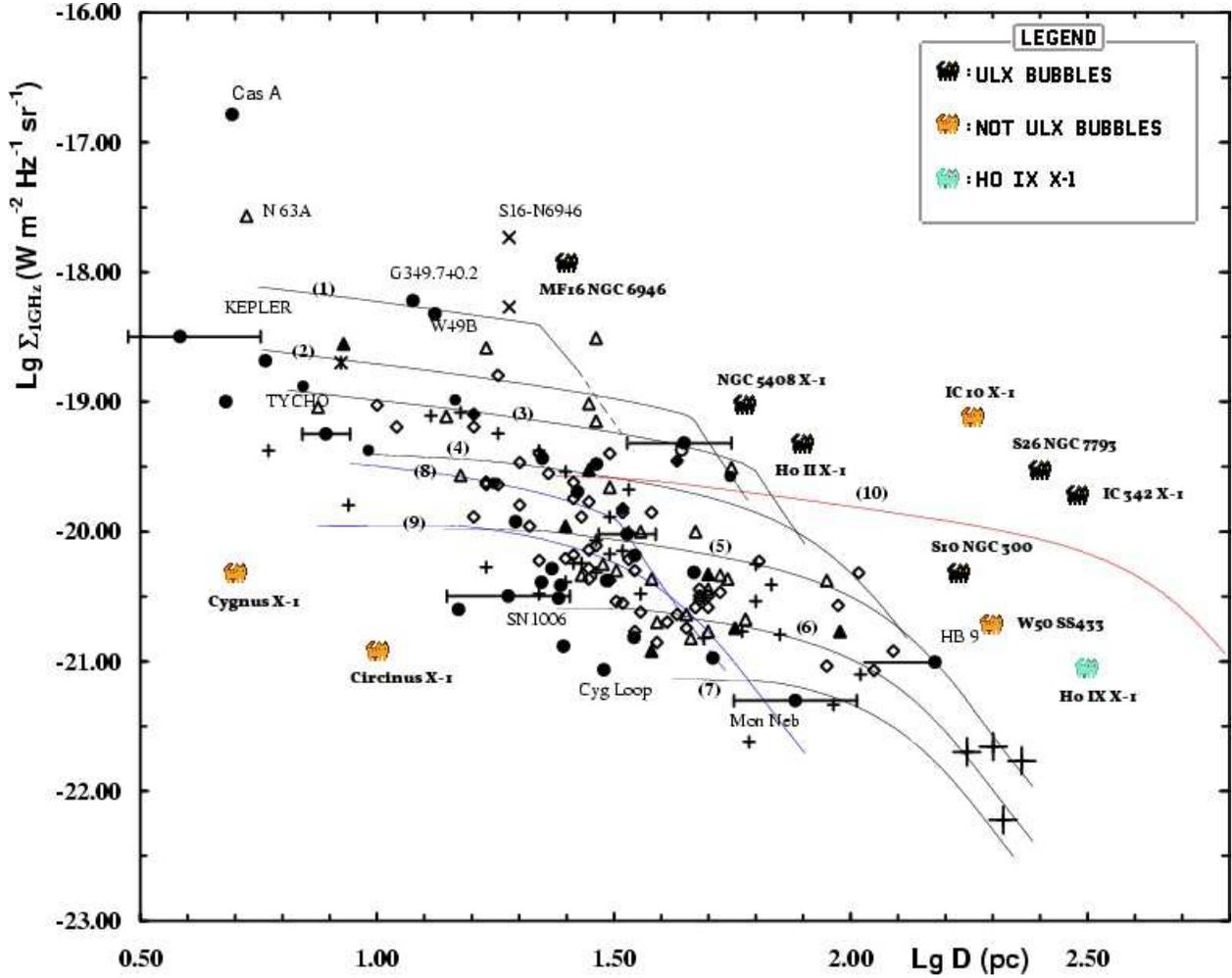}
\caption{Surface brightness-diameter relation ($\Sigma-\mathrm{D}$) from \citet{asv06}. The models are  for typical SNRs of energy 10$^{50}$~erg (blue), 10$^{51}$~erg (black) for decreasing gas densities, and 5$\times10^{52}$~erg (red).}
\label{sbright}
\end{figure*}

If the nebula is a supernova remnant, its high energy content suggests a hypernova. The structure appears to be more plerionic than shell-like. We can use the (1~GHz) surface brightness diameter relation ($\Sigma-\mathrm{D}$) from \citet{asv06}  to compare with known Galactic and some extragalactic SNRs. We use our band-to-band spectral index of -0.56 to extrapolate to 1 GHz and obtain a surface brightness 9.0$\times10^{-22}$ W m$^{-2}$ Hz$^{-1}$ sr$^{-1}$. In Figure \ref{sbright} we show the evolutionary models and SNRs from \citet{asv06}. We overplot Ho IX X-1 and other known ULX radio bubbles plus bubbles similar to W50/SS~433. Two of our objects (W50/SS~433 and S26 NGC 7793) were already plotted in \citet{asv06} but we used more recent published data as shown in Table \ref{asvarof}. While many of the ULX bubbles are known to have energies in excess of the higher model plotted here (5$\times10^{52}$~erg), even some of the lower energy cases such as W50/SS~433 do not fit these SNR models well. 

\begin{deluxetable*}{ccccc}
\tablecaption{Data table corresponding to Figure \ref{sbright}.\label{asvarof}}
\tablehead{
\colhead{Object Name} & \colhead{Spectral Index} & \colhead{Log $\sum_{1GHz}$} & \colhead{Log Diameter} & \colhead{Reference} \\
\colhead{} & \colhead{} & \colhead{$W m^{-2} Hz^{-1} sr^{-1}$} & \colhead{pc} & \colhead{}}
\startdata
 \text{Ho IX X-1} & -0.56 & -21.0 & 2.5 & This work\\
 \text{S26 NGC 7793} & -0.70 & -19.5 & 2.4 & \citet{sor10} \\
 \text{W50 SS433} & -0.48 & -20.7 & 2.3 & \citet{dub98} \\
 \text{IC 342 X-1} & -0.50 & -19.7 & 2.5 & \citet{cseh12} \\
 \text{Ho II X-1} & -0.53 & -19.3 & 1.9 & \citet{cseh12} \\
 \text{NGC 5408 X-1} &-0.80 & -19.0 & 1.8 & \citet{cseh12} \\
 \text{MF16 NGC 6946} & -0.50 & -17.9 & 1.4& \citet{van94} \\
 \text{S10 NGC 300} & -0.40 & -20.3 & 2.2 & \citet{urq19} \\
 \text{IC 10 X-1} & -0.41 & -19.1 & 2.3 & \citet{loz07} \\
 \text{Circinus X-1} & -0.50 & -20.9 & 1.0 & \citet{tud06} \\
 \text{Cygnus X-1} & -0.50 & -20.3 & 0.7 & \citet{gal05} \\
\enddata
\tablecomments{We use -0.5 as a substitute for missing spectral indexes.}
\end{deluxetable*}

We can estimate the initial explosion energy of the supernova following \citet{cio88} using the expansion velocity $>20$~km~s$^{-1}$ and an electron preshock density $n = 5$~cm$^{-3}$, based on the H$_\beta$ line \citep{abo08}. We obtain $E > 4.7\times10^{53}$ erg, a very powerful hypernova. \citet{ram06} estimated that an energy of only about $6\times10^{51}$~erg could have been released if multiple supernovae exploded in the OB association located within the bubble so this scenario is not very likely.

\subsection{Scenario 2: Jet/Wind-Blown Type  SS~433 Bubble} 

We did not detect a jet in our observations but it is possible that it exists, it could be fainter than we expected or it could be variable. \citet{mil05} found a radio nebula associated with Ho~II X-1 and later \citet{cseh12} detected jets with a deeper VLA observation. The best example of a jet-blown bubble is the microquasar SS~433, with its associated bubble W50 \citep{dub98}. Another good example is S~26 in NGC~7793 \citep{sor10}, which has a similar size and energy with Ho~IX~X-1 and also Ho~II X-1 \cite{cseh14, cseh15}. Additional evidence for a variable jet in Ho~IX~X-1 was recently reported by \citet{lau19}. The authors analyzed Spitzer data taken before and after the data used in \citet{dud16} and found no other detection. They suggest that \citet{dud16} observed an outburst and ruled out a circumbinary disk as the source of IR, therefore strengthening the variable jet interpretation. We intend to perform deeper or repeated VLA monitoring to look for a jet in Ho~IX~X-1.

In addition to a jet, bubble nebulae can also be inflated by winds, which are expected to be powerful in super-Eddington sources such as SS~433 \citep{beg06}. In fact there is evidence for a radio outflow in near the equatorial plane of SS~433, possibly driven by a disk wind \citep{par99, blu01}. It is therefore likely that both jets and winds are inflating the Ho~IX X-1 bubble and we are now considering both mechanisms.

We can estimate the radiative power following \citet{cseh12} based on the H$\beta$ line in \citet{abo08} who found shock velocities in the range 20--100~km~s$^{-1}$. We obtain P$_{rad} = (1.6-4.2)\times 10^{39}$ erg s$^{-1}$. The total power P$_{tot}$ = 77/27 P$_{rad} = (0.5 - 1.2) \times10^{40}$ erg s$^{-1}$. This is consistent with similar estimates in \citet{abo07, abo08}.

To get the total energy we multiply by the characteristic age (Sedov age) of the bubble. This can be estimated  using the shock velocities above and the bubble radius (200~pc). We obtain $t =1.2-5.9$~Myr and the total energy E$_{tot} = (4.4 - 8.6)\times10^{53}$ erg.

Following \citet{cseh12} and \citet{koe08} we can estimate the radio jet power from the total power estimated above. For a typical jet we obtain a jet power greater than $1.1 \times10 ^{34}$~erg s$^{-1}$, which is a factor of at least 20 times higher than the upper limit of our radio observations. This luminosity is consistent with what we expected from the extrapolation of the infrared excess we found \citep{dud16}, suggesting once more that if there is a radio jet responsible for inflating the bubble, it is probably variable. 

Let us now consider if the winds alone could produce the radio  bubble in Ho~IX~X-1 ULX. Using a long \textit{Chandra} observation of Ho~IX~X-1 to look for X-ray outflows \citet{wal13} did not detect any strong winds but they could still be present. \citet{siv17} simulated wind inflated ULX bubbles assuming typical physical properties for the ULX and the surrounding medium and estimated the emission at different wavelengths, including radio and X-ray. In their fiducial model the radiative phase happens after 0.5~Myr at a radius of 100~pc. At 5~GHz the luminosity starts at about $10^{34}$ erg s$^{-1}$ and increases to $10^{35}$ after about 0.3 Myr. Our estimated 5~GHz luminosity is $1.4\times10^{34}$ erg s$^{-1}$. Given the Sedov age of a few Myr, the above measured luminosity is slightly lower than the model of \citet{siv17}. Recently, \citet{sat19} attempted to detect the Ho~IX~X-1 bubble in X-rays using the Chandra X-ray Observatory. By stacking multiple X-ray images they were not able to detect the bubble to a limiting luminosity of $2\times10^{36}$ erg s$^{-1}$. This is also consistent with \citet{siv17}, as their fiducial model predicts X-ray luminosities peaking at $2\times10^{35}$ erg s$^{-1}$ but dropping by a factor of 10 after 0.3 Myr. 

It is finally probable that both jet and winds play a role just as in SS~433 and therefore the estimates of \citet{siv17} are a lower limit to the expected radio emission. In a similar way the estimated jet power in the only-jet scenario would be overestimated. A combination jet/winds as in SS~433 would actually better explain the data because the upper limit for a jet is still 20 times lower than our estimates required to inflate the bubble. If the jet is variable as the current data suggest, this represents the average flux needed, so actual variations are expected to be even higher. In SS~433 variations are only of the order of a factor of a few above base level \citep[see][and references therein]{bro18}. Same with Ho~II X-1, measured variations so far are less than a factor of 10 \citep{cseh15}.

\section{Conclusions} \label{section: Conclusions}
We performed C- and X-band VLA observations to look for a jet in Ho~IX~X-1 ULX. We did not detect a jet with an upper limit of 6.6 $\mu$Jy in the C band, which is lower than the expected flux extrapolated from the infrared excess detected in our previous paper. However, the infrared flux is variable and therefore the jet could be also variable as in Ho~II~X-1, for example. 

However, we discovered extended radio emission slightly smaller in size than the optical bubble in both radio bands, increasing the small sample of known ULX radio bubbles. We calculate the minimum energy assuming synchrotron emission ($4.4\times10^{50}$~erg) and the lifetime of electrons (70.4~Myr). We proposed two scenarios for the radio nebula based on our radio observations and previous optical studies. For an SNR interpretation we find that it requires a very high explosion energy, greater than $4.7\times10^{53}$~erg and therefore suggests a hypernova. We compare the $\Sigma-\mathrm{D}$ relation with known Galactic
and other ULX bubbles and find that Ho~IX bubble is similar to other ULX radio bubbles but not described well by SNR models. This is also true for radio bubbles associated to other microquasars similar to W50 (SS~433).

Although we did not directly detect a jet in our VLA observations and winds were not yet observed in Ho~IX~X-1, given the above considerations we prefer the jet/wind-blown bubble interpretation similar to the microquasar SS~433. We estimate that the total energy is greater than $4.4\times10^{53}$~erg, with a characteristic timescale of a few million years. These are much larger than SS~433 but similar to other ULX bubbles. We measure a 5~GHz luminosity of $1.4\times10^{34}$~erg~s$^{-1}$, while wind models suggest $10 ^{35}$~erg s$^{-1}$. On the other hand a jet power greater than $1.1 \times10 ^{34}$~erg s$^{-1}$ is required to inflate the bubble, which is a factor of at least 20 times higher than the upper limit of our radio observations. Therefore a a combination jet/winds as in SS~433 could better explain the data.

\acknowledgments
We thank the referee for very helpful suggestions that greatly improved our paper.  C.B. thanks Abdul Asvarov for the permission to use Figure \ref{sbright}. The National Radio Astronomy Observatory is a facility of the National Science Foundation operated under cooperative agreement by Associated Universities, Inc. This research made use of Astropy, a community-developed core Python package for Astronomy \citep{2013A&A...558A..33A,2018AJ....156..123A} and the package Common Astronomy Software Applications \citep[\textsc{casa},][]{McMullin+07}.

\facilities{VLA}

\software{Astropy \citep{2013A&A...558A..33A,2018AJ....156..123A}, \textsc{casa} \citep{McMullin+07} }



\end{document}